\RequirePackage{fix-cm}
\documentclass[smallcondensed]{svjour3}     
\usepackage{amsmath}
\usepackage{amssymb}
\usepackage{graphicx}
\usepackage{epstopdf}
\usepackage{subcaption}
\usepackage{floatrow}
\usepackage{subfloat}
 \usepackage{array}
 \usepackage{multirow}
\usepackage{algorithm}
\usepackage{algpseudocode}
\usepackage{authblk}
\usepackage{soul}
\usepackage[titletoc,title]{appendix}

\soulregister\cite7
\soulregister\ref7
\begin{document}

\title{Boosting quantum annealer performance via sample persistence 
}


\author{Hamed Karimi${^{1,2}}$         \and
         Gili Rosenberg${^1}$ 
}

\authorrunning{H. Karimi, G. Rosenberg} 

\institute{Hamed Karimi \at
 		{hamed.karimi@1qbit.com} \\
           \and
           Gili Rosenberg \at
	     {gili.rosenberg@1qbit.com} \\
	    \and
	     $^1$             1QB Information Technologies (1QBit), 458-550 Burrard Street, Vancouver, British Columbia \,V6C 2B5, Canada \\ \\
	     $^2$		     Department of Computer Science, University of British Columbia, 201-2366 Main Mall,
Vancouver, British Columbia \,V6T 1Z4, Canada	\\
}
%

\maketitle

\begin{abstract}
We propose a novel method for reducing the number of variables in quadratic unconstrained binary optimization problems, using a quantum annealer (or any sampler) to fix the value of a large portion of the variables to values that have a high probability of being optimal. The resulting problems are usually much easier for the quantum annealer to solve, due to their being smaller and consisting of disconnected components. This approach significantly increases the success rate and number of observations of the best known energy value in samples obtained from the quantum annealer, when compared with calling the quantum annealer without using it, even when using fewer annealing cycles. Use of the method results in a considerable improvement in success metrics even for problems with high-precision couplers and biases, which are more challenging for the quantum annealer to solve. The results are further enhanced by applying the method iteratively and combining it with classical pre-processing. We present results for both Chimera graph-structured problems and embedded problems from a real-world application.
\end{abstract}

\section{Previous work}

\subsection{Quantum annealing}
\label{sec:quantum_annealing}

Quantum annealers are now commercially available \cite{johnson2011quantum,bunyk2014architectural}. Manufactured by \mbox{D-Wave} Systems, they are designed to heuristically find low-energy states of an Ising model of the type
\begin{align}
H = \sum_i h_i s_i + \sum_{\langle i, j \rangle} J_{ij} s_i s_j
\label{equation:ising}
\end{align}
on a Chimera graph \cite{bunyk2014architectural}. The quantum annealer used in this study was the D-Wave 2X--SYS4, which has 1152 qubits, with a qubit yield of over $95\%$ (the remainder of the qubits are inoperable).\footnote{The chip at our disposal has 1100 active qubits, a working temperature of 26 $\pm$5 mK, and a minimum annealing time of 20 $\mu$s.} Since numerous well known NP-hard problems have known mappings to an Ising model \cite{lucas2014ising}, it is hoped that many intractable problems can be solved more efficiently using quantum annealing. 

There is evidence that quantum effects, namely entanglement and finite-range multi-qubit tunnelling, do indeed occur on these quantum annealing chips \cite{lanting2014entanglement,boixo2016computational,boixo2014evidence,denchev2015computational}. It has been suggested that quantum annealers have an advantage over classical optimizers due to quantum tunnelling, which an optimizer can exploit to search the solution space of an optimization problem by passing through energy barriers. For certain problem classes, this might provide a quantum speedup \cite{ray1989sherrington,finnila1994quantum,kadowaki1998quantum,santoro2002theory,battaglia2005optimization,heim2015quantum,muthukrishnan2015tunneling}. For this reason, there has been much recent interest in benchmarking quantum annealers against classical solvers, although conclusive evidence of quantum speedup has not been shown to date \cite{boixo2014evidence,denchev2015computational,mcgeoch2013experimental,katzgraber2014glassy,king2015performance,king2015benchmarking,mandra2016strengths,ronnow2014defining,hen2015probing,katzgraber2015seeking}. 

In order to solve a problem with the quantum annealer, the user must provide the biases $h$ and the couplings $J$ in Eq. \ref{equation:ising}. These parameters are affected by perturbations of different types, which are generally referred to as intrinsic control errors (ICE). Due to these errors, which are roughly \mbox{$2\%$--$4\%$} of the range of the parameters, high-precision problems are challenging for the D-Wave 2X, motivating post-processing and error correction efforts \cite{mishra2016performance,pudenz2014error,pudenz2015quantum,vinci2015quantum,perdomo2016determination,pastawski2015error,lechner2015quantum}, as well as our own efforts in this research. 

The hardware graph of the D-Wave 2X quantum annealer is a Chimera graph \cite{bunyk2014architectural} consisting of a $12\times12$ lattice of complete bipartite units of 8 qubits with sparse connectivity between the unit cells. Most real-world optimization problems are unlikely to have a corresponding Ising model with this structure. To allow the solving of problems with different structure, we use an \emph{embedding}, which is a mapping from each logical variable to one or more physical qubits, referred to as a \emph{chain}. We connect a chain of qubits with a strong ferromagnetic (negative $J_{ij}$) coupling, thereby penalizing solutions whose values are different. 

When obtaining a sample from the quantum annealer for an embedded problem, it is possible that not all of the qubits in a chain will have the same state. In this case the values of the corresponding logical variables are commonly determined by either majority voting or a process referred to as energy minimization \cite{vinci2015quantum}. In a version of the latter that we employed in this study, we assigned an effective field to each broken chain, chose the chain with the strongest effective field, set its state to be opposite to that of the direction of the field (to minimize the energy), updated the effective fields, and repeated the process until no broken chains remained. This is a quick method of local error correction. We remark that finding an embedding for a graph is formally known as graph-minor embedding, an NP-hard problem \cite{choi2008minor,choi2011minor} commonly solved by heuristic methods, although for some classes of problems deterministic embeddings can be found \cite{zaribafiyan2016systematic}. 

Other hybrid quantum-classical methods have been devised, including using the quantum annealer to set variables iteratively to the lowest-energy configuration in patches until convergence \cite{rosenberg2015building}, and using the annealer to guide a classical solver \cite{tran2016hybrid}. 

\subsection{Classical pre-processing}
\label{sec:classical_preprocessing}

Within the combinatorial optimization community, there is a significant body of work on methods for solving quadratic unconstrained binary optimization (QUBO) problems of the form
\begin{align}
&\min x^TQx \\
&\, \mbox{s.t.} \, x \in \{0,1\}^N, \nonumber 
\end{align} 
where $Q$ is a real $N \times N$ matrix. QUBO problems can easily be mapped to the above Ising model (with the addition of a constant term). Of interest are classical methods for pre-processing QUBO problems (see Tavares \cite{tavares2008new} for a review), in particular \emph{persistency}  \cite{hammer1984roof,Boros2006PreprocessingOU}, whereby some of the variables are fixed to their values in the optimum. When a variable is fixed, its value is set and the entries in $Q$ are updated accordingly. In the Ising-based approach, fixing a qubit adds a constant term to the energy and shifts the biases ($h$ values) of the neighbouring qubits. 

An example of classical pre-processing is fixing variables within a heuristic algorithm by maintaining a reference set of elite solutions (typically obtained by performing a local search), and finding the variables that are often set to the same value, the idea being that they are likely to be set to that same value in the optimum \cite{wang2011effective,wang2013backbone,zhang2004configuration}. The \emph{backbone} of a problem is defined as the variables and corresponding values which are the same in all optima \cite{monasson1999determining}. This notion has also been extended to problems with a non-degenerate optimum \cite{roma2013backbone}. These techniques, as well as our method, can be described as finding an \emph{approximate backbone} \cite{jiang2009backbone}.

The closest existing work to our method is Chardaire et al. \cite{chardaire1995thermostatistical}, which uses the term \emph{thermostatistical persistency} to describe variables whose values can be fixed in simulated annealing, based on their values remaining constant as the temperature is decreased. Our method, which we refer to as \emph{sample persistence}, relies on fixing variables whose value is the same in multiple solutions in a sample obtained from a sampler. We alternate between using classical pre-processing and applying our method using a quantum annealer as the underlying sampler. We find that using both methods together leads to a significant boost in the number of variables that are fixed, with a high level of confidence that they are fixed to their value in an optimal solution. Below, we use the Ising representation of this problem exclusively. We remark that, due to the one-to-one mapping between Ising and QUBO problems, our method could be described equivalently using a QUBO representation.

The layout of this paper is as follows: our proposed method is presented in Section~\ref{sec:method}, numerical results are presented in Section~\ref{sec:results}, suggestions for parameter choice are presented in Section~\ref{sec:parameter_choice} and discussed in Section~\ref{sec:discussion}, and concluding remarks are given in Section~\ref{sec:conclusions}. 

\section{Method}
\label{sec:method}

In this section, we explain the details of our proposed method and its refinements for various cases, as well as the physical intuition behind the method. In Section~\ref{sec:basic_method}, we introduce the basic variable-reduction method, and in Section~\ref{sec:iterative_method}, we present the iterative method, which builds upon the basic method. In Section~\ref{sec:physical_intuition}, we explain our physical intuition on why and when this method is expected to succeed. In Section~\ref{sec:method_zero_h}, we discuss challenges of the zero bias case ($h=0$) and how we mitigated them within the framework of our method. Finally, in Section~\ref{sec:method_embedded}, we discuss how we applied our method to embedded problems. 

\subsection{Sample persistence variable reduction (SPVAR)}
\label{sec:basic_method}

The basic idea behind our method is that, for a given problem, we obtain a sample from the quantum annealer, find the variables that have exactly the same value across the entire sample, and fix them to that value. A quantum annealer is a heuristic solver, and a typical call to the machine involves at least hundreds, if not thousands, of quantum annealing cycles, resulting in a sample of low-energy solutions. For a general random Ising problem, with $J$ and $h$ chosen randomly, we observed that many spins (variables) maintain their state (value) in the sample obtained from the quantum annealer. The sample being obtained from low-energy solutions motivated our conjecture (and working assumption) that those variables might also maintain their values in the ground state(s) of the system. \\

We improved on the basic idea in multiple ways:
\begin{itemize}
\item Instead of obtaining a sample in one call to the quantum annealer, we obtained a better sample by performing multiple calls, and applied a random gauge to each. It has been shown that quantum annealers perform better when random gauges\footnote{A gauge, in this context, implies multiplying each spin operator by $\pm 1$.} are applied to problems, since this mitigates errors in the quantum hardware \cite{perdomo2015performance}.
\item Trimming the sample. We gave more weight to lower-energy solutions by ignoring higher-energy solutions of the sample, based on a given \textit{elite\textunderscore threshold} percentile. An alternative, which we leave for future study, would be to give higher weights to lower-energy solutions with some given weight function.
\item Relaxing the fixing criterion. In our basic method, we fix a variable if it maintains its state in exactly all of the solutions of the sample. We relaxed this criterion by introducing a \textit{fixing\textunderscore threshold} parameter, such that variables are only fixed if their mean absolute value (taken across the trimmed sample) is larger than the value of this parameter.
\end{itemize}

\begin{algorithm}[!htbp]
\begin{algorithmic}
\Require Ising problem ($J$, $h$), \textit{sample\textunderscore size}, \textit{sampler}, \textit{fixing\textunderscore threshold}, \textit{elite\textunderscore threshold} 
\State Obtain a sample of \textit{sample\textunderscore size} from the \textit{sampler}
\State Narrow down the solutions to the \textit{elite\textunderscore threshold} percentile
\State Find the mean value of each variable in all solutions
\State Fix the variables for which the mean absolute value is larger than \textit{fixing\textunderscore threshold}
\State Update $J$ and $h$ \\
\Return $J$, $h$, and a mapping from the fixed variables to the values to which they were fixed
\end{algorithmic}
\caption{SPVAR}
\label{algorithm:method}
\end{algorithm}

Algorithm \ref{algorithm:method}, \emph{SPVAR}, summarizes the steps of our improved basic method. The parameters of the algorithm are described in Appendix~\ref{appendix:parameters}. Fixing variables in this way can drastically reduce the size of the effective problem to be solved. However, if even one variable is fixed to a value contrary to the value of that variable in all of the optima, it becomes impossible to find an optimum by solving the reduced problem. To check whether this occurs, we solved each problem before and after fixing the variables using a heuristic solver with a long time-out, such that it would have a high likelihood of finding an optimum. If all of the variables were fixed correctly, the energy of the optimum before fixing the variables would be guaranteed to be equal to the energy of the optimum after fixing them. We tuned the parameters so that we fixed variables such that they took their value in one of the optima with high confidence.

We remark that setting \textit{fixing\textunderscore threshold} to 1.0 guarantees that, if an optimum was found by the quantum annealer, then no variables will be fixed contrary to their value in at least one of the optima. This can be shown by contradiction. Assume one variable is fixed contrary to its value in all of the optima. This would only occur if that variable's value was set to the contrary value in all of the solutions in the sample. However, an optimum was among the solutions in the sample, so at least one solution in the sample has that variable set to a different value (i.e., to the value in the optimum that was found), leading to a contradiction.

Decreasing \textit{fixing\textunderscore threshold} allows one to fix more variables, since the fixing criterion is relaxed, and decreasing \textit{elite\textunderscore threshold} focuses on fewer solutions, again fixing more variables. In both cases, the increase in the number of fixed variables comes with an increased risk of fixing variables contrary to their optimum value. Changing these parameters allows the user to optimize SPVAR for the class of problems being optimized, as well as based on the user's preference. For example, if solving the problem faster (i.e., using a smaller sample size) is a priority, the above parameters can be set to low values. See Appendix~\ref{appendix:thresholds} for a more detailed discussion of the thresholds.

We observed that, after applying SPVAR, the graph structure of the reduced problems becomes simpler than that of the original problem. Classical techniques can benefit from this, for example, from having a smaller treewidth and nodes of degree one. We return to this observation, and present related results, in Section~\ref{sec:results_decomposition}.

\subsection{Iterative sample persistence variable reduction (ISPVAR)}
\label{sec:iterative_method}

For maximum effect, SPVAR can be applied repeatedly \textit{num\textunderscore steps} times, fixing increasingly more variables. In order to increase the effectiveness of SPVAR, we also used a classical pre-processing method that uses roof duality and weak persistency \cite{Boros2006PreprocessingOU} to fix more variables,\footnote{We used the function \textit{fix\textunderscore variables} in D-Wave Systems' SAPI 2.3.1, which is the solver API used to access the quantum annealer \cite{vlad2016fix_variabes}.} after applying SPVAR.

In addition to those variables that maintain their value in the sample, there are other variables that do not have this property but are perfectly correlated with each other and change their value coherently. We fixed additional variables by exploiting these correlations. After each step of SPVAR, we calculated the pair-wise correlations of all of the variables, based on the \textit{correlation\textunderscore elite\textunderscore threshold} lowest-energy solutions in the sample. We then created a graph in which the nodes were the variables and the edge weights were the correlation coefficients, disregarding all correlation coefficients with an absolute value of less than \textit{correlation\textunderscore threshold}, and found the connected components in this graph. If the call to the classical pre-processor fixed at least one variable in any of the connected components, we fixed the entire component (unless it contained a frustrated loop). 

The resulting problems were significantly smaller (see Section~\ref{sec:results_decomposition}) and easier for the quantum annealer to solve (see Sections~\ref{sec:results_chimera}--\ref{sec:results_ott}). Algorithm \ref{algorithm:iterative_method}, \emph{ISPVAR}, summarizes the iterative algorithm. The parameters of the algorithm are described in Appendix~\ref{appendix:parameters}.
\begin{algorithm}[!htbp]
\begin{algorithmic}
\Require Ising problem ($J$, $h$), \textit{sample\textunderscore size}, \textit{sampler}, \textit{fixing\textunderscore threshold}, \textit{elite\textunderscore threshold}, \textit{num\textunderscore steps}
\For{each step of \textit{num\textunderscore steps}}
\State Apply SPVAR to find modified $J$, $h$
\State [Optional] Apply classical pre-processing to find modified $J$, $h$
\State [Optional] Fix variables via correlations to find modified $J$, $h$
\EndFor \\
\Return $J$, $h$, and a mapping from the fixed variables to the values to which they were fixed
\end{algorithmic}
\caption{ISPVAR}
\label{algorithm:iterative_method}
\end{algorithm}
%

\subsection{Physical intuition}
\label{sec:physical_intuition}

In the limit of very large local fields (i.e., biases), much larger than the couplings, it is obvious that, in order to find the ground state of the system, we need only to consider the local fields and can ignore the couplings. Similarly, for finite local fields, Zintchenko et al. \cite{zintchenko2015local} showed that it is possible to find the ground state configuration of a given spin by solving the problem restricted to the local neighbourhood of that spin. 
The authors considered spin glass problems given by Eq. \ref{equation:ising}, where $J_{ij}$ are chosen from a normal distribution with a mean of zero and a standard deviation of one, and $h_i$'s chosen each with a standard deviation of $\sigma_h$. 

The size of the relevant neighbourhood $l$ depends on the strength of the local fields $h$. In particular, Zintchenko et al. showed that for spin glass problems on a square lattice $l \propto \sigma_h^{-0.8}$. The length scale $l$ corresponds to the correlation length of the system, and correlations between spins with a distance larger than this length decay exponentially. As a result, by considering a cluster of spins around a given spin $s_0$ of a size larger than $l$,  $s_0$ is uncorrelated to the boundary spins of the cluster. By solving the problem for the spins in the cluster with an arbitrary configuration on the boundary, we can find the ground state configuration of $s_0$.

It is expected that, at finite fields and low energies, due to finite correlations, these clusters define a larger effective spin which experiences an effective field and possibly some small coupling with other clusters. As a result, each cluster will align with its effective local field with high probability. Some of these clusters will experience larger effective fields compared to the others and will, therefore, be more robust under perturbations. A good low-energy sampler can detect these clusters. This could explain the existence of spins that maintain their value in the low-energy samples obtained from the quantum annealer. The low-energy excitations of the system can be described by the effective interaction between these clusters. Flipping the clusters, which results in the spins in each cluster flipping coherently with other spins in the same cluster, justifies the additional fixing of variables via correlation, as explained in the previous section.

\begin{figure}[!ht] 
   \centering
    \includegraphics[width=0.8\linewidth]{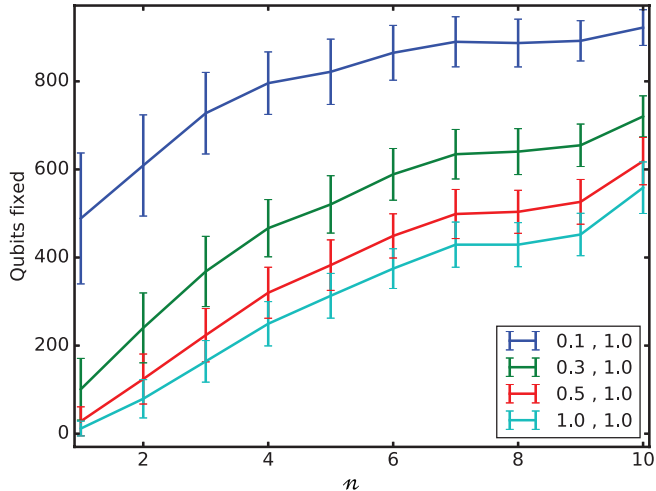} 
    \caption{Number of fixed qubits versus the range of $h$, for $J$ in $U_5$, for different pairs of parameters \textit{elite\textunderscore threshold} and \textit{fixing\textunderscore threshold} (respectively). SPVAR was applied using 2500 annealing cycles. The couplers were chosen uniformly at random over the set $\{-5, -4,\dots, 4, 5\}$, and the biases were chosen from the set $\{-n,\dots, n\}$, increasing $n$ from $1$ to $10$, which has the effect of changing the standard deviation of the bias $\sigma_h$. }
    \label{fig:J_const}
\end{figure}

Following this argument, we expect that detecting these clusters and fixing spins will become more difficult as the correlation length increases, in the limit of smaller local fields. Therefore, we conjectured that SPVAR would be challenged in the limit of $h \rightarrow 0$. In order to test this conjecture, we chose the couplers uniformly at random over the set $\{-5, -4,\dots, 4, 5\}$, but chose $h_i$ from the set $\{-n,\dots, n\}$,\footnote{The value zero was excluded for the couplers but not for the biases, and we use this convention throughout the paper.} increasing $n$ from $1$ to $10$. This has the effect of changing the standard deviation of the bias $\sigma_h$. The expected result was that increasing $\sigma_h$ will shorten the correlation length, and hence more variables should have been fixed by SPVAR. Figure~\ref{fig:J_const} presents the results of this experiment, which validate our intuition.

For the case of $h = 0$, the system has $\mathcal{Z}_2$ symmetry $\vec{s} \rightarrow -\vec{s}$. In this case, the sample might contain many solutions, along with their degenerate pairs, and SPVAR in this form will not work.  We present some modifications to deal with this situation in the following section. A simple approach to take is to simply break the $\mathcal{Z}_2$ symmetry by choosing a random spin and forcing its value to $+1$ or $-1$. With this approach, we were able to observe the correlation between the randomly chosen spin and the other spins.

\begin{figure}[!ht] 
   \centering
    \includegraphics[width=0.8\linewidth]{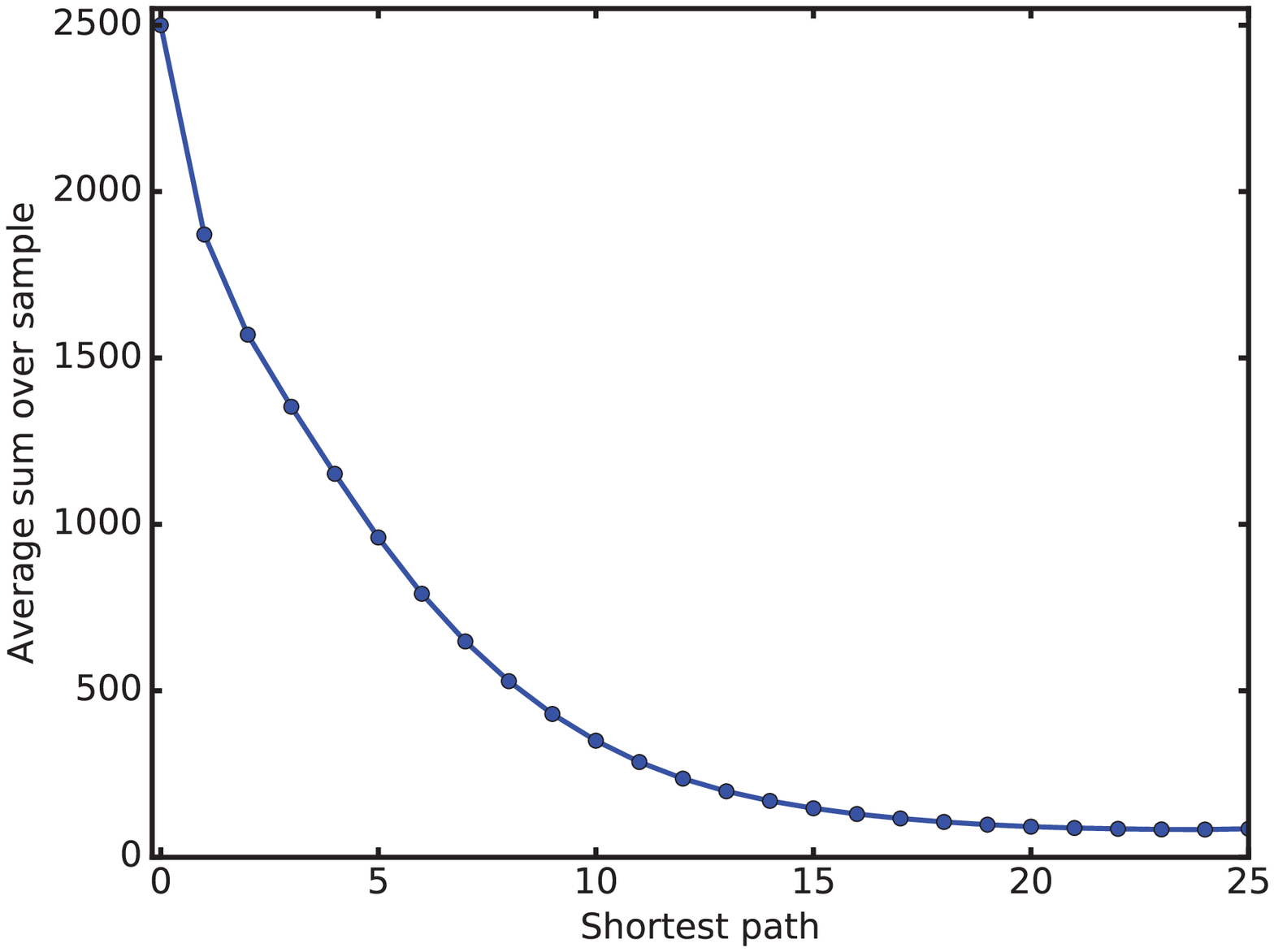}
    \caption{Average correlation between a fixed reference qubit and other qubits as a function of the shortest path. The couplers were chosen uniformly at random over the set $\{-5, -4,\dots, 4, 5\}$ and the biases were set to zero. For each problem, we obtained a sample from the quantum annealer, chose all of the solutions such that a reference qubit's state was `up', summed the states of all of the qubits, and averaged over all of the qubits that had the same length of shortest path.}
    \label{fig:correlation_length_study} 
\end{figure}

For this experiment, we created twenty Chimera graph-structured problems, based on the hardware graph of the D-Wave 2X (which has 1100 active qubits), with $h=0$ and $J$ chosen uniformly at random over the set \{$-5, -4,\dots, 4, 5$\}. For each problem, we first obtained a sample from the quantum annealer. We then iterated over the qubits, selecting one as our reference qubit. For each reference qubit, we chose all of the solutions such that this qubit's state was `up', summed the states of all of the qubits, and saved the results. Finally, we averaged over all of the qubits that had the same length of shortest path for any reference
qubit. See Figure~\ref{fig:correlation_length_study} for the average of the sum over the solutions (averaged over the problems and the reference qubits) as a function of the length of the shortest path.

%

\subsection{Special considerations for $h=0$ problems}
\label{sec:method_zero_h}

If all of the biases are zero, the system is time reversal invariant, leading to a two-fold degeneracy of all states. Therefore, we expect that the sample will contain many degenerate solution pairs. Since all variables would have an expected value of zero in the sample, this would lead to SPVAR fixing zero variables. For this reason, a modification was necessary in this case.

Without loss of generality, we fixed one variable of our choice to `up', thereby breaking the $\mathcal{Z}_2$ symmetry. We then iterated over the solutions in the sample, and for any solutions in which that variable's value was set to `down', we multiplied the entire solution by negative one, thereby recovering the degenerate pair for that solution. The result was that the expected state of each variable was no longer zero, and SPVAR was able to fix a non-zero number of variables.

There is freedom to choose any variable in the system to be fixed in the initial step. This motivated the idea that we should fix the variable that is the most correlated with other variables, which would allow us to fix those variables as well. As in Section~\ref{sec:basic_method}, we found the connected components in the correlation graph. In the initial step, we chose any variable from the largest connected component, and fixed all of the variables that were connected to it to their respective values. Using this method, in the initial pre-fixing step, we were able to fix a number of variables equal to the size of the largest connected component, instead of just one. 

In our experiments, we noted that the variance in the number of fixed variables was larger for the zero bias problems than for the non-zero bias problems. In particular, in the zero bias case, in some problems a large number of variables was fixed, and in some others a small number of variables was fixed. This motivated the question of how to decide, in advance, whether a particular problem could benefit from the modified ISPVAR. Katzgraber et al.~\cite{katzgraber2015seeking} found that problems for which the distribution of the spin glass order parameter is bimodal are easier, in principle. We also observed a correlation between the bimodality of the distribution of overlaps in the low-energy states and the number of fixed qubits. For this reason, we suggest using the overlap of low-energy states in a small sample obtained from a quantum annealer as a heuristic for predicting whether ISPVAR will be useful for particular zero bias problems. 

\subsection{Methods for embedded problems}
\label{sec:method_embedded}

Most Ising (or QUBO) problems that result from real applications do not have a Chimera graph structure, so they require a graph-minor embedding problem to be solved first, yielding a non-trivial mapping from variables (logical variables) to qubits (physical qubits). In order to use ISPVAR to solve such embedded problems, we devised two algorithms. The first, which we refer to as the \emph{logical method}, involved obtaining a sample from the quantum annealer at each step and converting it to a sample in the logical space by fixing the broken chains using an energy minimization method (see Section~\ref{sec:quantum_annealing}). ISPVAR was then applied in the logical space, fixing logical variables.

An alternative method, which we refer to as the \emph{physical method}, instead relied on embedding the problem and then fixing the states of the physical qubits at each step. In addition to fixing variables as in ISPVAR, we added two other fixing criteria. The first stipulated that if more than a fraction \textit{majority\textunderscore length\textunderscore threshold} of the qubits in a chain were fixed to the same state, then the remaining qubits in that chain were also fixed to that state (since it was assumed that that is the state of the corresponding logical variable in the optimum). For the second criterion, a relaxed fixing threshold \textit{chain\textunderscore fixing\textunderscore threshold} (typically 0.80--0.95) was imposed, and the qubits to be fixed were identified as in Section~\ref{sec:basic_method}, but only fixed if that would lead to fixing the whole chain. If fixing these qubits in a chain of greater length than \mbox{\textit{absolute\textunderscore min\textunderscore length}} (typically 3) resulted in all of the qubits being fixed, then they were actually fixed; otherwise, they were ignored.  

For the second criterion, a relaxed fixing threshold \textit{chain\textunderscore fixing\textunderscore threshold} (typically 0.80--0.95) was imposed, and the qubits to be fixed were identified as in Section~\ref{sec:basic_method}, but only actually fixed if that would lead to a fixed whole chain of a length greater than \mbox{\textit{absolute\textunderscore min\textunderscore length}} (typically 3).

\section{Results}
\label{sec:results}

This section is organized as follows. In Section~\ref{sec:results_procedure}, we describe the procedure we used for benchmarking ISPVAR. In Section~\ref{sec:results_chimera}, we present our results for Chimera graph-structured problems. In Section~\ref{sec:results_ott}, we present results for a case study on embedded problems, namely optimal trading trajectory problems. In Section~\ref{sec:results_decomposition}, we show the connected component sizes in the effective problems after applying our method, and in Section~\ref{sec:results_postprocessing}, we study the effect of applying a post-processing step to the quantum annealer's sample prior to applying our method. 

\subsection{Benchmarking procedure}
\label{sec:results_procedure}

We defined problem sets in which the biases and couplers were chosen uniformly from a given set of integers, for example, $U_2 \in \{-2,-1,0,1,2\}$ (0 was excluded for the couplers but included for the biases) and a Sidon set \mbox{$S_{28} \in \{-28, -19, -13, -8, 8, 13, 19, 28\}$}. For each of the problem sets, we generated 100 random instances. To get a baseline result, we used the quantum annealer to solve each problem with 50,000 annealing cycles (using five random gauges with 10,000 cycles each). We then applied ISPVAR on each problem, with $num\textunderscore steps=4$ and 2500 annealing cycles per step, finally using the quantum annealer to solve each reduced problem with 10,000 cycles (for a total of 20,000 cycles).\footnote{The reduced problems often consist of multiple connected components (see Section~\ref{sec:results_decomposition}). We took advantage of this fact when evaluating the energy values for the states in each sample.} This allowed us to compare solving a problem set with the quantum annealer alone versus solving the problem set by first applying ISPVAR. 

As solution metrics, we used the number of occurrences of the best known energy value, the energy residual, and the success rate. The number of occurrences is the mean number of occurrences of the best known energy value for each sample for which at least one such solution was found. The energy residual is the difference in energy between the best solution found and the best known solution. The success rate is the percentage of problem instances for which the best known energy was found. An often-used benchmarking practice is to measure the number of annealing cycles required for 99\% confidence of seeing the best known energy value at least once. However, in practice, this measurement can be impractical for hard problems or high-precision problems, since it requires a very large number of annealing cycles, and measuring the number of occurrences provides a proxy that is easier to measure. We also calculated the number of problem instances for which the lowest-energy solution found by our method, used in conjunction with with the quantum annealer, was at least as good as the lowest-energy solution found by the quantum annealer alone.  

In order to determine the success rate of the quantum annealer, we solved each problem instance with a heuristic with a long time-out, which gave a high level of confidence of finding the optimum (but not a guarantee). For the Chimera graph-structured problems, we used an implementation of the Hamze--de Freitas--Selby (HFS) algorithm \cite{hamze2004fields,selby2013github,selby2014efficient}, and for the embedded problems, we used an implementation of multi-start tabu 1-opt search \cite{glover1989tabu,glover1990tabu} with 1000 repetitions per problem instance. ISPVAR's fixing success rate was defined as the percentage of problem instances for which ISPVAR fixed variables to values which appeared in at least one best known solution. This was verified by finding the best known solution both before and after fixing variables. If our method succeeded, the global optimum would be preserved; if it failed, then the energy value of the optimum would be higher after fixing the variables than before fixing them. 
 
\subsection{Chimera graph-structured problems}
\label{sec:results_chimera}

Results for the number of fixed variables in each step for the Chimera graph-structured problem sets, for $h=0$ and $h \neq 0$, are presented in Table~\ref{table:chimera_multi_step_fixed}. For each step, we show a breakdown of the number of variables fixed, so that one can identify how many variables were fixed by SPVAR combined with variables fixed based on correlation, and how many were fixed due to classical pre-processing (see Section~\ref{sec:classical_preprocessing}), as well as the total number of variables fixed. Table~\ref{table:chimera_multi_step_success} compares the success rate of the quantum annealer alone with the success rate of ISPVAR and the quantum annealer used together, and shows the ISPVAR fixing success rate. 
\begin{table}[!htb]
\begin{tabular}{c|l|rr|rr|rr|rr|rr|r}
\hline
& \multirow{2}{*}{Set}    &    \multicolumn{2}{c|}{0}    &    \multicolumn{2}{c|}{1}       &     \multicolumn{2}{c|}{2}       &    \multicolumn{2}{c|}{3}     &     \multicolumn{2}{c|}{4}     &   \multirow{2}{*}{Total} \\
\cline{3-12}
&    &    m &    c &     m &     c &    m &    c &    m &    c &    m &    c &   \\
\hline
\parbox[t]{2mm}{\multirow{5}{*}{\rotatebox[origin=c]{90}{$h\neq 0$}}}
& $U_2$     &  --- & 11 & 517 & 168 &  69 & 48 & 52 & 33 &  45 & 20 &   964 \\
& $U_5$    &  --- &  3 & 520 & 138 &  95 & 47 & 64 & 33 &  55 & 21 &   976 \\
& $U_{10}$    &  --- &  3 & 503 & 125 & 101 & 50 & 69 & 29 &  65 & 23 &   967 \\
& $S_{28}$    &  --- &  0 & 550 &  97 &  92 & 38 & 62 & 28 &  62 & 25 &   954 \\
& $U_{100}$   &  --- &  2 & 508 & 122 &  85 & 44 & 55 & 31 &  52 & 22&   920 \\
\hline
\parbox[t]{2mm}{\multirow{5}{*}{\rotatebox[origin=c]{90}{$h = 0$}}}
& $U_2$   & 47 &  9 &   9 &   3 &  28 &  6 & 62 & 10 & 199 & 24 &   397 \\
& $U_5$   & 43 &  6 &  13 &   3 &  24 &  5 & 51 &  7 & 131 & 13 &   297 \\
& $U_{10}$  & 39 &  5 &  11 &   2 &  32 &  5 & 46 &  7 &  89 &  9 &   246 \\
& $S_{28}$  & 47 &  5 &  17 &   3 &  25 &  3 & 36 &  5 &  76 &  8 &   225 \\
& $U_{100}$ & 34 &  5 &   8 &   2 &  11 &  2 & 14 &  3 &  22 &  4 &   104 \\
\hline
\end{tabular}

\caption{Mean number of fixed variables for Chimera graph-structured problems, by problem set and by step. Step 0 is a pre-step in which only the classical pre-processing was run. For the zero bias ($h=0$) case, the pre-step also included fixing the largest connected component, as described in Section~\ref{sec:method_zero_h}. For each step of the four steps of ISPVAR, the mean number of of variables fixed by SPVAR and correlations (`m') and the classical pre-processing (`c') are shown. The final column shows the mean total number of fixed variables.}
\label{table:chimera_multi_step_fixed}
\end{table}
\begin{table}[!htb]
\begin{tabular}{c|l|rrr|rrr|rr}
\hline
  &  \multirow{2}{*}{Set}      & \multicolumn{3}{c|}{Quantum annealer} & \multicolumn{3}{c|}{ISPVAR} &  \multirow{2}{*}{Better}  & \multirow{2}{*}{Fix} \\
  &    &   Success &    Residual &   Freq &   Success &    Residual &    Freq & &  \\
\hline
\parbox[t]{2mm}{\multirow{5}{*}{\rotatebox[origin=c]{90}{$h \neq 0$}}}
& $U_2$     &     92 &       0.16 & 285 &     99 &       0.02 & 4831 &   100 & 99 \\
& $U_5$     &     27 &       2.92 &   5 &     91 &       0.24 & 1726 &   100 & 95 \\
& $U_{10}$  &   2 &      10.76 &   2 &     94 &       0.36 & 1127 &   100 & 99 \\
& $S_{28}$ &      0 &      29.84 &   0 &     82 &       1.88 &  852 &   100 & 95 \\
 & $U_{100}$   &      0 &     123.46 &   0 &     77 &      11.22 &  470 &   100 & 92 \\
\hline
\parbox[t]{2mm}{\multirow{5}{*}{\rotatebox[origin=c]{90}{$h=0$}}}

& $U_2$   &    57 &       1.10 &  19 &     73 &       0.56 & 1842 &    93 & 99 \\
& $U_5$   &     1 &       7.92 &   1 &     20 &       5.30 &  31 &    83 & 95 \\
& $U_{10}$  &      1 &      18.36 &   1 &     10 &      13.34 &  407 &    84 & 93 \\
& $S_{28}$  &      0 &      42.68 &   0 &     11 &      35.56 &  850 &    69 & 87 \\
& $U_{100}$  &      0 &     218.60 &   0 &      1 &     205.74 &    8 &    62 & 88 \\
\hline
\end{tabular}
\caption{Success metrics by problem set for the quantum annealer with and without ISPVAR, for Chimera graph-structured problems. `Success' is the percentage of problems (out of 100) for which the best known energy value was observed at least once. `Residual' is the mean energy difference between the lowest energy found and the best known solution. `Freq' is the mean number of occurrences of the best known solution's energy value in all solutions obtained from the quantum annealer, averaged over the problem instances for which the best known solution's energy value was seen at least once. The total number of annealing cycles was 50,000 for `Quantum annealer' and 20,000 for `ISPVAR' (2500 at each of the four steps of ISPVAR and 10,000 afterwards) for the finite field ($h \neq 0$) problems and 40,000 for the zero bias ($h=0$) problems (2500 at each of the four steps of ISPVAR, and 30,000 afterwards). `Better' shows the percentage of problems for which the lowest-energy solution found by ISPVAR was equal to or better than the lowest-energy solution found by the quantum annealer. The final column, `Fix', shows the percentage of problems for which ISPVAR fixed variables such that their values occur in a best known solution of the original problem.}
\label{table:chimera_multi_step_success}
\end{table}

For the first step of our method, the parameters \textit{fixing\textunderscore threshold} and {\textit{elite\textunderscore threshold} were tuned to 1.0 and 0.3, respectively, by generating a problem set with 100 problem instances in $U_5$, and scanning different parameter values. The parameters for the remaining three steps were set based on trial and error, to 0.2, 0.15, and 0.1 for \textit{elite\textunderscore threshold} and to 1.0 for all steps for \textit{fixing\textunderscore threshold} (to preserve the optimum, as described above). The reason we decreased \textit{fixing\textunderscore threshold} for the later steps is that, as the method progresses, increasingly more qubits are fixed and the problems become simpler, so we become increasingly confident that the quantum annealer will solve the problem. The correlation graph parameters \textit{correlation\textunderscore threshold} and \textit{correlation\textunderscore elite\textunderscore threshold} were set to 1.0 and 0.4, respectively, for all steps. These might not be the optimal parameters---selecting optimal values is left for future work. 

\subsection{Embedded problems -- case study}
\label{sec:results_ott}

A challenge in studying the performance of our method on embedded problems is the large range of problems available, as well as the necessity to control the parameters, given the objective of studying the dependence of the method's performance on the precision of the biases and couplers. In order to perform a case study, we obtained an adjacency matrix for a particular problem on which we generated random problem instances. 

The problem we studied was an optimal trading trajectory (OTT) problem \cite{rosenberg2016solving}, which is a multi-period portfolio optimization problem. In the OTT problem, the objective is to maximize the future returns of discrete financial assets, given forecast returns and risk, and taking into account transaction costs. We chose the largest problem instance that could be embedded on the chip available to us, which had eight assets, five time steps, and a bit depth of two, giving a total of 80 logical variables, which required 872 qubits and a longest chain of 14. 

We used the embedding and adjacency matrix of the original problem, but chose random biases and couplers. This preserved the structure of the original problem, while allowing us to control the precision of the couplers and biases. To embed the problems, we first scaled them down to the range $[-0.5,+0.5]$ and then used the \textit{embed\textunderscore problem} function in D-Wave Systems' SAPI 2.3.1 (such that the intra-chain couplers were set to $-1$). 

Results for the number of fixed variables in each step for the OTT problem sets are presented in Table~\ref{table:ott_multi_step_fixed}, for both the logical and physical methods (see Section~\ref{sec:method_embedded}). For each step, we show the total number of variables fixed due to SPVAR, the correlations, and the classical pre-processing (see Section~\ref{sec:classical_preprocessing}), as well as the mean total number of fixed variables. Table~\ref{table:ott_multi_step_success} compares the success rate of the quantum annealer used on its own with the success rate of ISPVAR and the quantum annealer used together, and shows ISPVAR's fixing success rate. 
\begin{table}[!htb]
\begin{tabular}{c|l|rrrrr|r}
\hline
& Set   &     \multicolumn{1}{c}{1} &      \multicolumn{1}{c}{2} &      \multicolumn{1}{c}{3} &     \multicolumn{1}{c}{4} &     \multicolumn{1}{c|}{c} &   Total \\
\hline
\parbox[t]{2mm}{\multirow{4}{*}{\rotatebox[origin=c]{90}{Logical}}} 
& $U_2$    & 4 & 14 & 11 & 11 & 3 &   44 \\
& $U_5$    & 4 & 11 & 11 & 13 & 3 &   42 \\
& $U_{10}$   & 5 & 12 & 11 &  10 & 2 &   39 \\
& $U_{100}$  & 3 &  8 &  8 &  7 & 4 &   30 \\
\hline
\parbox[t]{2mm}{\multirow{4}{*}{\rotatebox[origin=c]{90}{Physical}}} 
& $U_2$    & 3 & 16 & 10 &  8 & 4 &   41 \\
& $U_5$    & 2 & 14 & 11 &  7 & 4 &   38 \\
& $U_{10}$   & 3 & 15 &  9 &  6 & 3 &   35 \\
& $U_{100}$  & 2 & 12 &  9 &  6 & 5 &   33 \\
\hline
\end{tabular}
\caption{Mean number of fixed variables for embedded optimal trading trajectory (OTT) problems, by problem set, by step, and by method (logical or physical methods). For each of the four steps of ISPVAR, the mean number of fixed variables is shown due to fixing variables first via classical pre-processing and then via SPVAR and correlations. The last step consisted of fixing variables using only classical pre-processing. The final column shows the total mean number of fixed variables.}
\label{table:ott_multi_step_fixed}
\end{table}
\begin{table}[!htb]
\begin{tabular}{c|l|rrr|rrr|rr}
\hline
 &  \multirow{2}{*}{Set}  & \multicolumn{3}{c|}{Quantum annealer} & \multicolumn{3}{c|}{ISPVAR} & \multirow{2}{*}{Better} &  \multirow{2}{*}{Fix}  \\
& &   Success &   Residual &   Freq &   Success &   Residual &    Freq &  & \\
\hline
\parbox[t]{2mm}{\multirow{4}{*}{\rotatebox[origin=c]{90}{Logical}}} 
& $U_2$    &     60 &       0.67 &   6.40 &     79 &       0.38 & 2059 &    93 & 91 \\
& $U_5$   &     69 &       0.33 &  17.43 &     77 &       0.24 & 2085 &    86 & 93 \\
& $U_{10}$   &     71 &       0.24 &  13.48 &     77 &       0.23 & 2179 &    88 & 98 \\
& $U_{100}$  &     57 &       0.37 &   5.67 &     67 &       0.28 &  933 &    83 & 96 \\
\hline
\parbox[t]{2mm}{\multirow{4}{*}{\rotatebox[origin=c]{90}{Physical}}} 
& $U_2$    &     60 &       0.67 &   6.40 &     61 &       0.57 & 1105 &    86 & 81 \\
& $U_5$    &     69 &       0.33 &  17.43 &     66 &       0.32 & 1403 &    78 & 86 \\
& $U_{10}$   &     71 &       0.24 &  13.48 &     68 &       0.32 & 1246 &    81 & 89 \\
& $U_{100}$  &     57 &       0.37 &   5.67 &     65 &       0.34 &  730 &    84 & 90 \\
\hline
\end{tabular}
\caption{Success metrics by problem set for the quantum annealer with and without ISPVAR, for embedded OTT problems, for the logical and physical methods. See the caption of Table~\ref{table:chimera_multi_step_success} for a definition of the metrics.}
\label{table:ott_multi_step_success}
\end{table}

The parameter \textit{fixing\textunderscore threshold} was set to 1.0 for all steps in both methods. The parameter \textit{elite\textunderscore threshold} was set to 0.2, 0.2, 0.15, and 0.1 for the four steps of both methods. The correlation graph parameters \textit{correlation\textunderscore threshold} and \textit{correlation\textunderscore elite\textunderscore threshold} were set to 0.95 and 0.2, respectively, for all steps of the logical method and 1.0 and 0.2 for all steps of the physical method. The parameters \textit{chain\textunderscore fixing\textunderscore threshold}, \textit{chain\textunderscore elite\textunderscore threshold}, and \textit{majority\textunderscore length\textunderscore threshold} were set to 0.95, 0.2, and 0.51, respectively, for all steps of the physical method. 

\subsection{Problem decomposition}
\label{sec:results_decomposition}

We have already shown that ISPVAR reduces the number of variables. However, an additional benefit of the method is that the resulting (effective) problems tend to be decomposed into smaller connected components, many of them small enough to be solved quickly, even exhaustively. In addition, the larger-sized components have a simpler graph structure, such as much smaller treewidth, than the original problem, from which dynamic programming could benefit.  Figure~\ref{fig:component_sizes} shows a histogram of the sizes of the connected components for $U_5$, after four steps. 
\begin{figure}[!ht] 
   \centering
    \includegraphics[width=0.8\linewidth]{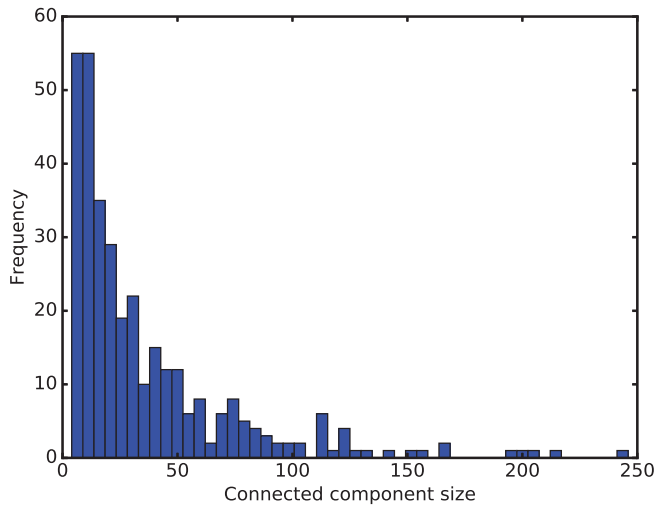} 
    \caption{Frequency of connected component sizes in the resulting 100 problems after 4 steps of ISPVAR for the $U_5$ problem set.}
    \label{fig:component_sizes}
\end{figure}
%

\subsection{Effect of post-processing}
\label{sec:results_postprocessing}

The performance of the quantum annealer can be improved by post-processing. We hypothesized that post-processing should improve the results of our method. In particular, it would increase the method's fixing success rate and improve the success metrics compared to using the quantum annealer alone (with post-processing). The results of running the quantum annealer with and without post-processing and with and without SPVAR are presented in Table~\ref{table:optimization}. Post-processing was done via the \emph{optimization} flag of the function \textit{solve\textunderscore ising} in D-Wave Systems' SAPI 2.3.1, which performs a local search on each of the solutions returned from the quantum annealer \cite{dwave2016sapi}. 
\begin{table}[ht]
\begin{tabular}{l|rrr|rrr|r}
\hline
\multirow{2}{*}{Post-processing} &  \multicolumn{3}{c|}{Quantum annealer} & \multicolumn{3}{c|}{SPVAR} & \multirow{2}{*}{Fix} \\
                                &   Success &    Residual &   Freq  & Success &    Residual &   Freq &\\
\hline
Without               &      0 & 123.46 &   0 &      1 &  96.16 &   1 & 99 \\
 With           &     60 &   9.24 &  50  &     87 &   4.20 & 458 & 100 \\
\hline
\end{tabular}\caption{Success metrics for the quantum annealer without post-processing, with post-processing, for the quantum annealer alone (`Quantum annealer') and with SPVAR, for problem set $U_{100}$. See the caption of Table~\ref{table:chimera_multi_step_fixed} for a definition of the columns.}
\label{table:optimization}
\end{table}


\section{Parameter choice}
\label{sec:parameter_choice}

We expect that the adjustable parameters in our method could be tuned based on the class of problems being optimized so as to yield the best results for that class of problems. In our study, we purposefully chose parameters such that they were not over-fit. For example, we did not choose different parameters for the high-precision problems compared to the low-precision problems, even though we observed that for the low-precision problems better results can be achieved by setting \textit{elite\textunderscore threshold} lower (since these problems are easier for the quantum annealer to solve). The objective here was to choose parameters that are a good all-around choice for general use of our method. In general, we have found that the method is quite robust with respect to the parameters, so it is not necessary nor is it advisable to fine-tune them. 

For practitioners interested in using our method, we offer some recommendations on parameter choice. First, setting \textit{fixing\textunderscore threshold} to a value of one is advisable, since that guarantees that results obtained after applying ISPVAR will not be worse, in principle, than the original sample (see Section~\ref{sec:basic_method} for an explanation). Decreasing the fixing threshold has the effect of increasing the number of variables that can be fixed. This same effect can also be achieved by modifying \emph{elite\textunderscore threshold}, which is what we recommend. In order to set \textit{elite\textunderscore threshold}, we suggest progressively increasing it until the fraction of variables fixed is 30\%--40\%. This is an inexpensive operation since, for each value of the threshold, the same original sample is re-evaluated. Setting \textit{elite\textunderscore threshold} too small is risky, since it can result in the elite sample being very small and lacking diversity. 

When using ISPVAR, we advise decreasing \textit{elite\textunderscore threshold} for each step, under the assumption that the resulting problems become easier and, hence, the sample becomes better. We observed that this gives better results than a constant \textit{elite\textunderscore threshold}. When choosing the number of steps, there is a tradeoff---having fewer steps leads to the fixing of fewer variables and a lower risk of fixing them incorrectly. Regarding the sample size to use for fixing variables, using a very small sample leads to a higher risk of fixing variable incorrectly, and using a very large sample is wasteful. 

\section{Discussion}
\label{sec:discussion}

For Chimera graph-structured problems, for the non-zero bias case ($h \neq 0$), Table~\ref{table:chimera_multi_step_fixed} shows that at least $84\%$ of the variables were fixed, on average, in four steps of ISPVAR. Notably, classical pre-processing used on its own was able to fix at most only $1\%$ of the variables. Table~\ref{table:chimera_multi_step_success} shows that, for all problem sets, the results with ISPVAR have significantly better success metrics, that is, higher success rates, lower mean-energy residuals, and a higher mean number of occurrences of the best known energy value. This is compared to using the quantum annealer alone, despite using less annealing cycles. In addition, the lowest-energy solution found with ISPVAR was always of an equal or lower energy value than the lowest-energy solution found by the quantum annealer alone. The percentage of problems for which variables were fixed incorrectly was low, less than $8\%$, and could be reduced further by choosing a higher \textit{elite\textunderscore threshold}. 

The zero bias problems (i.e., $h=0$) are known to be harder than the non-zero bias problems, but ISPVAR still fixed $10\%$--$36\%$ of the variables. The results with ISPVAR show improved success metrics, also fixing variables incorrectly for up to $13\%$ of the problems. For these problems, the mean number of variables fixed increased at every step, in contrast to the non-zero bias problems, where it decreased. The explanation for this is that, as variables are fixed, they induce non-zero biases on the neighbouring qubits, so that after each additional step of the method, we observe more qubits with non-zero bias, and the resulting problem becomes more like the non-zero bias problems. 

For the embedded OTT problems, Table~\ref{table:ott_multi_step_fixed} shows that $38\%$--$54\%$ of the variables were fixed by the logical method and $41\%$--$51\%$ were fixed by the physical method. In this case, the classical pre-processing was initially unable to fix any of the variables. Table~\ref{table:ott_multi_step_success} shows that, for all problem sets, the results with ISPVAR have significantly improved success metrics. In addition, the method's lowest-energy solution was as good as, or better than, the lowest-energy solution for the quantum annealer used alone, for the majority of the problems. 

We found that the logical method was better than the physical method, for the problem sets and specific parameters chosen, giving improved success metrics. In addition, the logical method fixed variables incorrectly less often than the physical method. We conjecture that this is due to the physical method being more sensitive to the quality of the embedding. In particular, the optimum of the embedded (i.e., physical) problem is not guaranteed to include a solution with unbroken chains. In principle, this could be guaranteed by increasing the strength of the intra-chain couplings, but the limited range of the biases and couplers in the quantum annealer limits our ability to do so. The physical method fixes variables with the objective of reaching the optimum of the embedded problem. However, if the embedded problem does not include an unbroken solution as one of the optima, the solution to which the physical method converges might not lead to the optimum of the logical problem. 

Our results show better success metrics for both Chimera graph-structured problems and embedded problems, even for high-precision problems that appear to be very difficult for the quantum annealer. Our intuition is that the quantum annealer is better at finding low-energy states than it is at finding optima. For difficult problems, for which the expected number of occurrences of the optimum would be very low, using only the lowest-energy solution found does not make full use of the information in the sample obtained from the quantum annealer. It appears that the sample contains additional information about the structure of the low-energy states, information that allows our method to boost the performance of the quantum annealer. 

In Section~\ref{sec:basic_method}, we described how to find correlation-based connected components. In principle, all of the variables in a connected component could be reduced to a single variable, by identification (i.e., by adding a strong coupling between them), such that they are either equal, or opposite, to that single variable, making full use of the information in the correlations. The issue with doing this for Chimera graph-structured problems is that the resulting problem may no longer be Chimera graph-structured (since new ferromagnetic couplers must be introduced), and as such might not fit on the chip (i.e., it would be impossible to find an embedding). In the case of embedded problems, the same issue arises, but in that case there is a more realistic possibility that an embedding could be found for the new problem. This is not guaranteed, however, and it would require an additional and possibly costly intermediate embedding step. Another possibility for utilizing our method is as a pre-processing step for a classical solver, as such solvers are typically not limited to a particular structure. 

Each generation of quantum annealers has exhibited a lower level of noise (i.e., ICE) than the previous one; therefore, future quantum annealers are likely to show further reductions. The results in Section~\ref{sec:results_postprocessing}, for using the quantum annealer with and without post-processing, give an indication that an improvement in this regard should also result in the improved performance of our method. These results (see Table~\ref{table:optimization}) show that the results for the quantum annealer used alone and those of our method were both drastically improved, with increased success rate, lower mean residual, and a higher mean number of occurrences of the best known solution. 

\section{Conclusions and future work}
\label{sec:conclusions}

We have presented a new method, ISPVAR, for the iterative quantum processing of QUBO problems, which yields a significant reduction in the number of variables in a problem. Consequently, it results in a pronounced increase in success rates and the number of occurrences of the best known energy value, when compared against a quantum annealer alone, even when using less annealing cycles. For this reason, at least for the problem sets studied herein, this evidence strongly suggests that this method is a better way to use the quantum annealer to greater potential, for both Chimera graph-structured problems and embedded problems. 

More-specific tuning of parameters for specific problem classes could yield even better results, and future research might uncover improvements to the method presented in this work. Future work could investigate the application of this method to other problem sets---native Chimera or embedded problems. The risk of fixing variables incorrectly suggests that it might be advisable to apply the method and solver multiple times (possibly in parallel), mitigating this risk. We leave a detailed scaling analysis of the success metrics, with and without the method, for future work. 

It has been suggested that the quantum annealer exhibits quantum speedup, compared to simulated annealing, when collective tunnelling occurs \cite{denchev2015computational}. It is possible that our method would benefit even from finite-range tunnelling, a hypothesis that we leave for future study. The method's performance and the fine-tuning of its parameters for other problem classes is another possible avenue for future study.

\begin{acknowledgements}
{ }The authors would like to thank Dominic Marchand, Pooya Ronagh, and Brad Woods for their insightful comments, Marko Bucyk for editing the manuscript, and Alex Selby for the use of his implementation of the Hamze--de Freitas--Selby (HFS) algorithm, available for public use on GitHub \cite{selby2013github}. This work was supported by 1QBit and Mitacs.

\noindent \textbf{Conflict of Interest} { }Hamed Karimi is an academic intern and Gili Rosenberg is an employee at 1QBit. 1QBit is focused on solving real-world problems using quantum computers. D-Wave Systems is a minority investor in 1QBit.
\end{acknowledgements}

\bibliographystyle{spmpsci}      

\bibliography{Boosting_QA_Performance}   

%
%

\clearpage
\begin{appendices}

\section{Dependence on thresholds}
\label{appendix:thresholds}

To illustrate the dependence of the number of fixed variables and SPVAR's fixing success rate on \textit{fixing\textunderscore threshold} and \textit{elite\textunderscore threshold}, we present results for two problem sets in Table~\ref{table:single_step_chimera_fixed} and Table~\ref{table:single_step_chimera_success}. The fixing success rate was defined as the percentage of problems for which the method fixed variables only to their optimal value. Detailed results for two choices of parameters, for the same two problem sets, are presented in Figure~\ref{fig:single_step_chimera}. In that experiment, SPVAR was applied on a quantum annealer sample obtained from 2500 annealing cycles. We defined the quantum annealer's residual as the energy difference between the quantum annealer's lowest-energy solution and the best known solution (found by a heuristic solver; see Section~\ref{sec:results} for more details), and the method's residual as the difference in energy between the best known solution before fixing variables and after fixing variables. 

We see that the quantum annealer was able to find the best known solution for most of the low-precision problems (top row), but unable to find the best known solution for almost all of the higher-precision problems (bottom row). The method's residual was almost always zero, indicating that the method almost never fixed variables incorrectly. In the few cases in which it did fix variables incorrectly, the effective problem (after fixing variables) still conserved the first or second excited states (the energy spacing was exactly two, due to the integer biases and couplers). We also note that for those problems, the method's residual was always lower than the quantum annealer's residual. The mean fraction of variables fixed was $58\%$--$70\%$, showing that SPVAR was able to fix most of the variables in these problems, and, as shown in Section~\ref{sec:results_chimera}, the iterative method presented (ISPVAR) was able to fix even more variables. 

\begin{figure}[!h] 
  \begin{subfigure}[b]{0.5\linewidth}
    \centering
    \caption{}
    \includegraphics[width=1.0\linewidth]{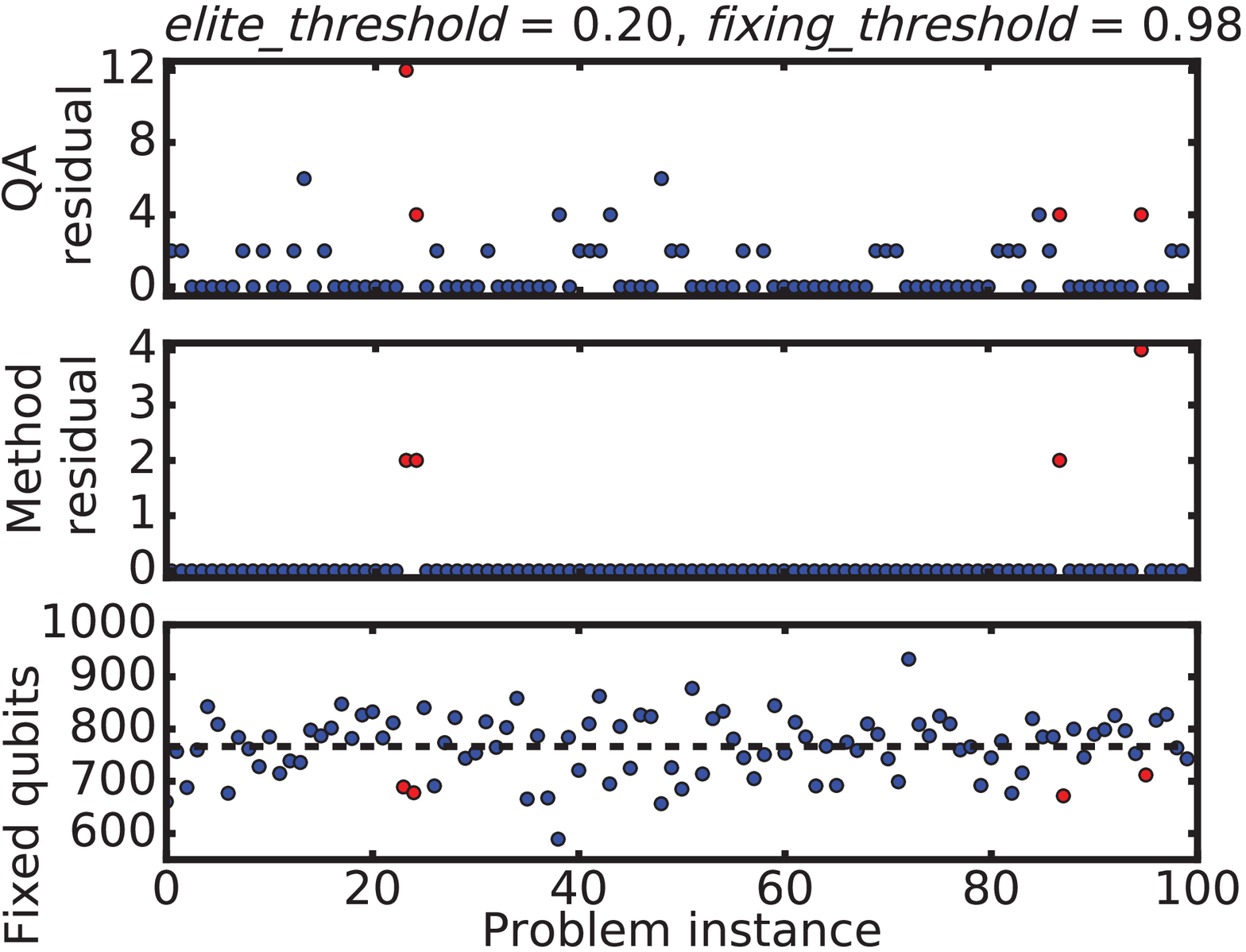} 
    \label{fig:single_step_chimera:a} 
  \end{subfigure}
  \begin{subfigure}[b]{0.5\linewidth}
    \centering
    \caption{}
    \includegraphics[width=1.0\linewidth]{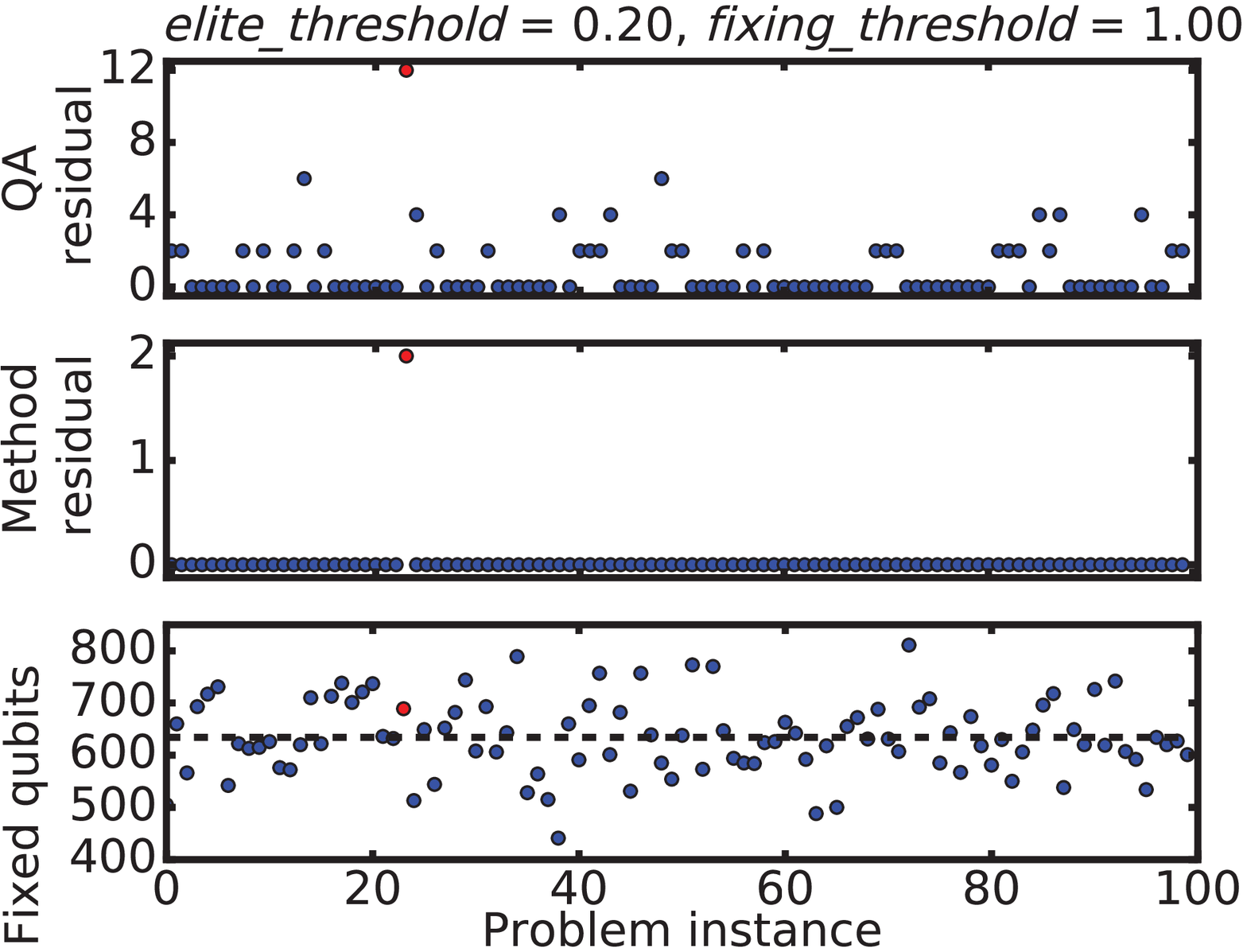} 
    \label{fig:single_step_chimera:b} 
  \end{subfigure} 
  \begin{subfigure}[b]{0.5\linewidth}
    \centering
        \caption{}
    \includegraphics[width=1.0\linewidth]{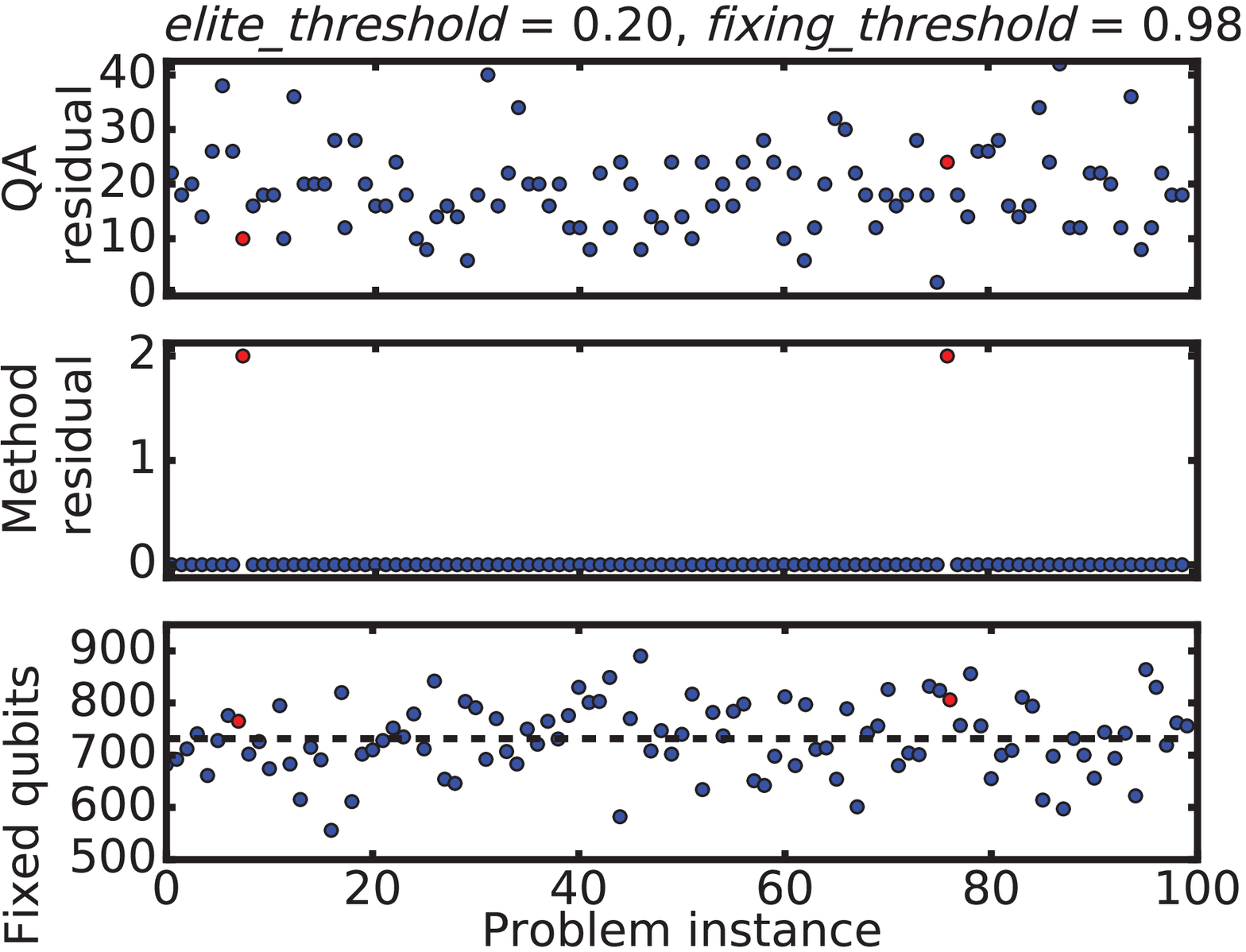} 
    \label{fig:single_step_chimera:c} 
  \end{subfigure}
  \begin{subfigure}[b]{0.5\linewidth}
    \centering
    \caption{}
    \includegraphics[width=1.0\linewidth]{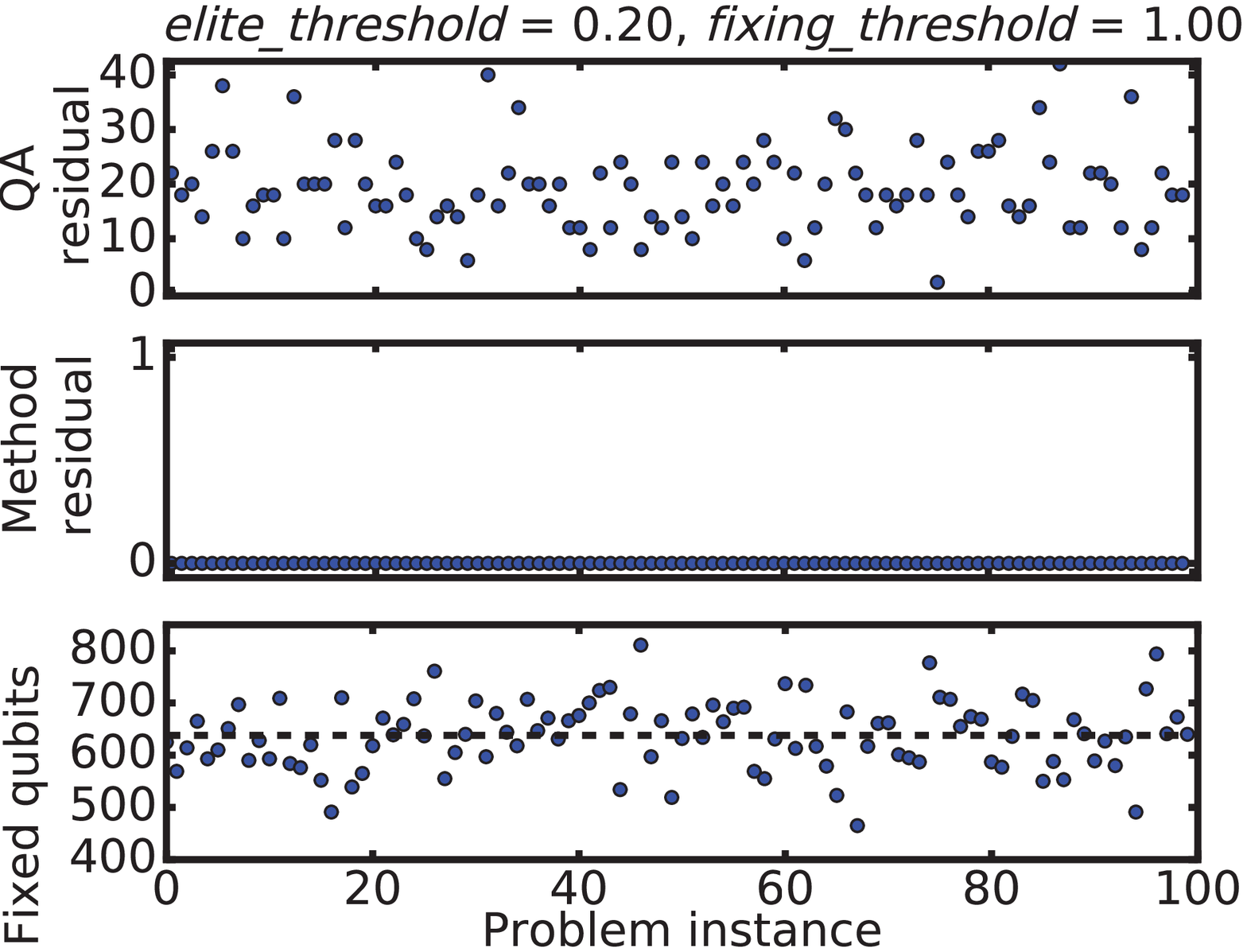} 
    \label{fig:single_step_chimera:d} 
  \end{subfigure} 
  \caption{Results for SPVAR on Chimera graph-structured problems. \textbf{(a)} and \textbf{(b)} present results for problems in $U_2$ ($J,h \in \{-2,-1,0,1,2\}$, excluding 0 for $J$), and \textbf{(c)} and \textbf{(d)} present results for problems in $U_{10}$. The parameters are indicated in the titles above each subfigure. The top panel of each subfigure shows the difference in energy between the quantum annealer's best solution and the best known solution, for each problem, obtained from 2500 annealing cycles using five random gauges. The middle panel shows the difference in energy between the best known solution before fixing variables and the best known solution after fixing variables (using the same 2500 cycles)---it should be zero if all variables fixed by SPVAR were fixed correctly. The bottom panel shows the number of fixed variables for each problem. A horizontal dashed line indicates the mean number of fixed variables. In all panels, the points that correspond to problems for which SPVAR fixed variables incorrectly are show in red, and the rest are shown in blue [colour online].}
  \label{fig:single_step_chimera} 
\end{figure}
\begin{table}[!h]
\centering
\begin{subtable}{.5\textwidth}
\centering
\caption{Mean number of fixed variables for $U_2$}
\begin{tabular}{r|rrrrr}
\hline
   E \textbackslash \,F &   0.95 &   0.97 &   0.98 &   0.99 &   1.00 \\
\hline
                      0.10 & 892 & 873 & 858 & 841 & 835 \\
                      0.20 & 828 & 794 & 766 & 718 & 634 \\
                      0.30 & 794 & 754 & 723 & 671 & 525 \\
                      0.40 & 769 & 729 & 697 & 639 & 455 \\
                      0.60 & 753 & 712 & 680 & 618 & 407 \\
                      0.60 & 746 & 704 & 670 & 606 & 384 \\
                      0.70 & 742 & 698 & 663 & 600 & 372 \\
                      0.80 & 740 & 696 & 659 & 596 & 365 \\
                      0.90 & 739 & 695 & 658 & 594 & 363 \\
                      1.00 & 739 & 694 & 661 & 594 & 362 \\
\hline
\end{tabular}
\end{subtable}
\vspace{0.5cm}\\
\begin{subtable}{.5\textwidth}
\centering
\caption{Mean number of fixed variables for $U_{10}$}
\begin{tabular}{r|rrrrr}
\hline
   E \textbackslash \,F &   0.95 &   0.97 &   0.98 &   0.99 &   1.00 \\
\hline
                      0.10 & 842 & 819 & 816 & 816 & 816 \\
                      0.20 & 798 & 758 & 732 & 686 & 638 \\
                      0.30 & 760 & 718 & 687 & 638 & 515 \\
                      0.40 & 733 & 689 & 656 & 603 & 427 \\
                      0.60 & 715 & 670 & 635 & 579 & 368 \\
                      0.60 & 705 & 656 & 620 & 564 & 333 \\
                      0.70 & 697 & 647 & 612 & 551 & 312 \\
                      0.80 & 693 & 643 & 607 & 548 & 300 \\
                      0.90 & 692 & 641 & 604 & 543 & 294 \\
                      1.00 & 690 & 640 & 605 & 541 & 290 \\
\hline
\end{tabular}
\end{subtable}
\caption{Mean number of fixed variables for different \textit{fixing\textunderscore threshold} (columns: `F') and \textit{elite\textunderscore threshold} (rows: `E') using SPVAR on Chimera graph-structured problems (with 1100 variables). Problems were chosen randomly from \textbf{(a)} $U_2$ ($J,h \in \{-2,-1,0,1,2\}$, excluding 0 for $J$) and \textbf{(b)} $U_{10}$ (defined similarly).}
\label{table:single_step_chimera_fixed}
\end{table}
\begin{table}[!h]
\centering
\begin{subtable}{.5\textwidth}
\centering
\caption{SPVAR fixing success rate for $U_2$}
\begin{tabular}{r|rrrrr}
\hline
   E \textbackslash \,F &   0.95 &   0.97 &   0.98 &   0.99 &   1.00 \\
\hline
                      0.10 &   0.91 &   0.93 &   0.94 &   0.96 &   0.96 \\
                      0.20 &   0.90 &   0.94 &   0.96 &   0.98 &   0.99 \\
                      0.30 &   0.85 &   0.93 &   0.96 &   0.97 &   0.99 \\
                      0.40 &   0.84 &   0.91 &   0.94 &   0.97 &   0.99 \\
                      0.60 &   0.83 &   0.90 &   0.93 &   0.95 &   0.99 \\
                      0.60 &   0.83 &   0.89 &   0.93 &   0.95 &   0.99 \\
                      0.70 &   0.83 &   0.89 &   0.93 &   0.95 &   0.99 \\
                      0.80 &   0.83 &   0.89 &   0.93 &   0.95 &   0.99 \\
                      0.90 &   0.83 &   0.89 &   0.93 &   0.95 &   0.99 \\
                      1.00 &   0.83 &   0.89 &   0.93 &   0.95 &   0.99 \\
\hline
\end{tabular}
\end{subtable}
\vspace{0.5cm} \\
\begin{subtable}{.5\textwidth}
\centering
\caption{SPVAR fixing success rate for $U_{10}$}
\begin{tabular}{r|rrrrr}
\hline
   E \textbackslash \,F &   0.95 &   0.97 &   0.98 &   0.99 &   1.00 \\
\hline
                      0.10 &   0.94 &   0.95 &   0.95 &   0.95 &   0.95 \\
                      0.20 &   0.88 &   0.95 &   0.98 &   0.99 &   1.00 \\
                      0.30 &   0.85 &   0.95 &   0.98 &   0.99 &   1.00 \\
                      0.40 &   0.83 &   0.93 &   0.99 &   0.99 &   1.00 \\
                      0.60 &   0.82 &   0.91 &   0.96 &   0.99 &   1.00 \\
                      0.60 &   0.81 &   0.92 &   0.97 &   0.99 &   1.00 \\
                      0.70 &   0.81 &   0.92 &   0.98 &   0.99 &   1.00 \\
                      0.80 &   0.81 &   0.91 &   0.98 &   0.99 &   1.00 \\
                      0.90 &   0.81 &   0.90 &   0.98 &   0.99 &   1.00 \\
                      1.00 &   0.81 &   0.90 &   0.98 &   0.99 &   1.00 \\
\hline
\end{tabular}
\end{subtable}
\caption{SPVAR fixing success rate for different \textit{fixing\textunderscore threshold} (columns: `F') and \textit{elite\textunderscore threshold} (rows: `E') on Chimera graph-structured problems. Problems were chosen randomly from \textbf{(a)} $U_2$ and \textbf{(b)} $U_{10}$ (see the caption of Table~\ref{table:single_step_chimera_fixed} for a definition of $U_2$ and $U_{10}$). The SPVAR fixing success rate is defined as the percentage of problem instances for which the method fixed variables to values which appeared in at least one best known solution, which 
was found by solving the problems with a heuristic method before and after fixing the variables.}
\label{table:single_step_chimera_success}
\end{table}

\clearpage
\section{Parameter description}
\label{appendix:parameters}

In Table~\ref{table:parameters}, we list the parameters used in the text and give a short description of each. 

\begin{table}[!h]
\centering
\begin{tabular}{c|p{8cm}}
\hline
	Parameter & Description \\
\hline
        \textit{sampler} & A sampler that returns a low-energy sample; in this work, it was always a quantum annealer \\
        \hline
        \textit{sample\textunderscore size} & The size of the sample used for fixing variables  \\
        \hline
        \textit{fixing\textunderscore threshold} & The fraction of the sample in which a variable must have the same value in order to be fixed \\
        \hline
        \textit{elite\textunderscore threshold} & The lowest-energy fraction of the sample which is used for fixing variables \\
        \hline
        \textit{num\textunderscore steps} & The number of steps to apply SPVAR in the iterative algorithm \mbox{ISPVAR} \\
        \hline
        \textit{correlation\textunderscore threshold} & The minimum absolute value of correlation between two variables for them to be considered correlated \\
        \hline
        \textit{correlation\textunderscore elite\textunderscore threshold} & The lowest-energy fraction of the sample which is used for fixing variables based on correlations \\
        \hline
        \textit{majority\textunderscore length\textunderscore threshold} & The fraction of physical variables that must agree in a chain in order to fix the entire chain (only for the physical method) \\
        \hline
        \textit{chain\textunderscore fixing\textunderscore threshold} & The fraction of the sample in which a variable must have the same value in order to be fixed---this is a relaxed fixing threshold, which only leads to fixing variables if fixing those variables would lead to an absolute majority, fixing the whole chain (only for the physical method)  \\
        \hline
        \textit{absolute\textunderscore min\textunderscore length} & The minimum length of a chain in order to apply  \textit{chain\textunderscore fixing\textunderscore threshold} (only for the physical method) \\
\hline
\end{tabular}
\caption{Parameters of SPVAR and ISPVAR}
\label{table:parameters}
\end{table}



\end{appendices}

\end{document}